\documentclass[12pt,preprint]{aastex}

\shorttitle{Propagation of the 2006 December 13 CME}

\shortauthors{Liu et al.}

\begin{document}

\title{A Comprehensive View of the 2006 December 13 CME: From the Sun
to Interplanetary Space}

\author{Y. Liu\altaffilmark{1,2}, J. G. Luhmann\altaffilmark{1},
R. M\"{u}ller-Mellin\altaffilmark{3}, P. C.
Schroeder\altaffilmark{1}, L. Wang\altaffilmark{1}, R. P.
Lin\altaffilmark{1}, S. D. Bale\altaffilmark{1}, Y.
Li\altaffilmark{1}, M. H. Acu\~{n}a\altaffilmark{4}, and J.-A.
Sauvaud\altaffilmark{5}}

\altaffiltext{1}{Space Sciences Laboratory, University of
California, Berkeley, CA 94720, USA; liuxying@ssl.berkeley.edu.}

\altaffiltext{2}{State Key Laboratory of Space Weather, Chinese
Academy of Sciences, Beijing 100080, China.}

\altaffiltext{3}{Institut f\"{u}r Experimentelle und Angewandte
Physik, Universit\"{a}t Kiel, Kiel, Germany.}

\altaffiltext{4}{NASA Goddard Space Flight Center, Greenbelt,
Maryland, USA.}

\altaffiltext{5}{Centre d'Etude Spatiale des Rayonnements, Centre
National de la Recherche Scientifique, Toulouse, France.}

\begin{abstract}
The biggest halo coronal mass ejection (CME) since the Halloween
storm in 2003, which occurred on 2006 December 13, is studied in
terms of its solar source and heliospheric consequences. The CME is
accompanied by an X3.4 flare, EUV dimmings and coronal waves. It
generated significant space weather effects such as an
interplanetary shock, radio bursts, major solar energetic particle
(SEP) events, and a magnetic cloud (MC) detected by a fleet of
spacecraft including STEREO, ACE, Wind and Ulysses. Reconstruction
of the MC with the Grad-Shafranov (GS) method yields an axis
orientation oblique to the flare ribbons. Observations of the SEP
intensities and anisotropies show that the particles can be trapped,
deflected and reaccelerated by the large-scale transient structures.
The CME-driven shock is observed at both the Earth and Ulysses when
they are separated by 74$^{\circ}$ in latitude and 117$^{\circ}$ in
longitude, the largest shock extent ever detected. The ejecta seems
missed at Ulysses. The shock arrival time at Ulysses is well
predicted by an MHD model which can propagate the 1 AU data outward.
The CME/shock is tracked remarkably well from the Sun all the way to
Ulysses by coronagraph images, type II frequency drift, in situ
measurements and the MHD model. These results reveal a technique
which combines MHD propagation of the solar wind and type II
emissions to predict the shock arrival time at the Earth, a
significant advance for space weather forecasting especially when in
situ data are available from the Solar Orbiter and Sentinels.
\end{abstract}


\section{Introduction}

Coronal mass ejections (CMEs) are the most spectacular eruptions in
the solar atmosphere and have been recognized as primary drivers of
interplanetary disturbances. They are called interplanetary CMEs
(ICMEs) when they move into the solar wind. Often associated with
CMEs and ICMEs are radio bursts, shock waves, solar energetic
particle (SEP) events, and prolonged southward magnetic field
components. A southward field component can reconnect with
geomagnetic fields and produce storms in the terrestrial environment
\citep[e.g.,][]{dungey61}. Understanding CMEs and characterizing
their interplanetary transport are crucial for space weather
forecasting but require coordinated multi-wavelength observations in
combination with in situ measurements.

The 2006 December 13 CME is the largest halo CME since the Halloween
storm which occurred in October - November 2003
\citep[e.g.,][]{gopalswamy05, richardson05, lario05}, given the
observed speeds of the CME and its forward shock, the time duration
of the ICME at 1 AU, the SEP intensities and the angular extent of
the shock (see \S 2 and \S 3). It is also the largest CME in the era
of the Solar TErrestrial RElations Observatory (STEREO) up to the
time of this writing. Different from the Halloween storm, this event
is relatively isolated from other CMEs, so contamination by or
mixing with other events is less pronounced; propagation into a
solar wind environment near solar minimum would also make
theoretical modeling easier. Accompanied by an X3.4 solar flare, the
CME evolved into a magnetic cloud (MC) and produced significant
space weather effects including SEP events, an interplanetary shock
and radio bursts detected by various instruments aboard a fleet of
spacecraft. Examining the evolution and propagation of this event
through the heliosphere would provide benchmark studies for CMEs,
associated phenomena and space weather.

The purpose of this work is to study the solar source and
heliospheric consequences of this CME in the frame of the Sun-Earth
connection. We combine EUV, coronagraph, radio, in situ particle,
plasma and magnetic field measurements with modeling efforts in an
attempt to give a comprehensive view of the event; particular
attention is paid to tracking the CME/shock all the way from the Sun
far into interplanetary space. We look at EUV and coronagraph images
in \S 2. Evolution of the CME in the heliosphere and its effects on
particle transport are investigated in \S 3. In \S 4, we combine
different data and demonstrate how the CME/shock propagation can be
tracked using coordinated observations and magnetohydrodynamic (MHD)
modeling. The results are summarized and discussed in \S 5.

\section{CME at the Sun}

We look at CME observations from the Large Angle Spectroscopic
Coronagraph (LASCO) and coronal observations from the
Extreme-ultraviolet Imaging Telescope (EIT) aboard the SOlar and
Heliospheric Observatory (SOHO). The Sun Earth Connection Coronal
and Heliospheric Investigation (SECCHI) of STEREO was not turned on
at the time of the CME. Figure~1 displays combined images as seen by
EIT and LASCO/C2 during the CME times. The CME starts with a strong
EUV brightening in the southwest quadrant at 02:30 UT. Around 02:54
UT, it forms a nearly complete halo; the EUV brightening is followed
by a dimming which quickly spreads into a diffusive area of the
solar disk. The CME moves further out around 03:06 UT and forms a
spectacular ring of dense material. The good timing between the
dimming and CME indicates that the reduced EUV brightness results
from removal of the coronal plasma due to the lift-off of the CME
\citep[e.g.,][]{thompson98, zarro99}. Depletion of the coronal
material by CMEs is possible if the associated magnetic field is
opened or stretched into interplanetary space as proposed by many
CME models. Note that a small CME precedes the big event as can be
seen in Figure~1 (left). Interactions of successive CMEs are thought
to affect SEP production \citep[e.g.,][]{gopalswamy04}.

Faint diffuse EUV brightenings are also seen and appear to be
propagation fronts of the dimming. These brightenings may represent
coronal waves propagating away from the active region. They were
first discovered by \cite{neupert89} but popularized in EIT
observations by \cite{thompson98}; for that reason they have been
referred to as ``EIT waves". The low cadence rate of EIT
observations (12 min for the 195 \AA\ band) does not allow an
accurate determination of the speed of the waves. The brightenings
moving toward the northeast hemisphere, however, seem to have a
constant speed: they travel a distance of $\sim$0.86 $R_{\odot}$
(solar radius) within 24 min from 02:24 UT to 02:48 UT and another
0.43 $R_{\odot}$ within 12 min from 02:48 UT to 03:00 UT (see
Figure~1). The speed is estimated to be about 420 km s$^{-1}$. The
CME speed projected on the sky is about 1774 km s$^{-1}$ as measured
along a position angle of 193$^{\circ}$ (counter clockwise from the
north; see CME identification and parameters at the LASCO CME
catalogue \url{http://cdaw.gsfc.nasa.gov}), significantly larger
than the EIT wave speed. It is not clear about the nature or origin
of the coronal waves, although an unambiguous correlation between
EIT waves and CMEs has been established
\citep[e.g.,][]{biesecker02}. An observed metric type II burst
starting at 02:27 UT, however, indicates that the present EIT wave
is likely a shock wave.

A closer look at the images also reveals a sharp edge all the way
around the CME front (see middle and right panels of Figure~1),
reminiscent of shock signature. Given the fast expansion of the CME
and a density at 1 AU comparable to the ambient solar wind (see
Figure~3), the CME density near the Sun must be much larger than the
ambient density; the density increase due to shock compression is at
most a factor of 4 of the background medium. The sheath region (a
transition layer between the CME front and shock) should thus have a
brightness weaker than the CME, consistent with the coronagraph
observations. Therefore, the sharp white-light feature is likely the
CME-driven shock. It is very rare to see the shock in white light,
especially for halo CMEs \citep[e.g.,][]{vourlidas03}. Relationship
between the EIT waves, metric type II burst and white-light shock
will be further investigated in a separate work.

The CME is accompanied by an X3.4 solar flare located in the active
region NOAA 10930 (S06$^{\circ}$W23$^{\circ}$). The flare seems to
be induced by a strong shear in the magnetic field associated with a
filament eruption, leading to two large ribbons which are twisted
but largely horizontal around the filament channel
\citep[e.g.,][]{zhang07, kosovichev07}. We will compare the
orientation of the filament channel with reconstruction of the
associated MC observed in situ \cite[see \S 3.1 and][]{li07}

\section{Interplanetary Consequences}

After the abrupt formation in the solar corona, the CME propagates
into the interplanetary medium and is observed in situ by STEREO,
ACE and Ulysses. We infer the ICME structure from in situ
measurements of plasma and magnetic field parameters combined with a
flux-rope reconstruction model. Connectivity of the ICME back to the
Sun is indicated by energetic particles which could be channeled,
constrained and reaccelerated by the transient structure.

\subsection{ICME at 1 AU}

STEREO observed the ICME after an exit from the terrestrial
magnetosheath. Figure~2 shows STEREO in situ measurements across the
event from the Solar Wind Electron Analyzer
\citep[SWEA;][]{sauvaud07} and the magnetometer
\citep[MAG;][]{acuna07} of the in situ measurements of particles and
CME transients investigation \citep[IMPACT;][]{luhmann07}. STEREO A
and B were not well separated, so they observed essentially the same
structure. The plasma parameters (e.g., density, velocity and
temperature) are not available from STEREO for that time period.
Bi-directional streaming electrons (BDEs) seem coincident with the
strong magnetic fields, indicative of closed field lines within the
event; rotation of the field (see the field elevation angle)
indicates an MC. The MC interval is determined from the BDEs but
also consistent with the reduced field variance and the rotation of
the field. The magnetic field has a significant negative (southward)
component which caused a major geomagnetic storm with
$D_{st}\sim-$190 nT. Interestingly, there is a current sheet
(indicated by the peak of the field elevation angle) within the MC,
which might be due to the passage of the comet McNaught through the
event (C. T. Russell, private communication). The magnetic field
trailing behind the MC has a roughly constant direction, presumably
stretched by the MC because of its large speed. A preceding shock,
as can be seen from simultaneous increases in the electron flux and
magnetic field strength, passed the spacecraft at 14:38 UT on
December 14. The transit time is about 36 hr from the Sun to the
Earth (assuming a launch time 02:30 UT on December 13), suggesting
an average speed of $\sim$1160 km s$^{-1}$. This speed is
significantly smaller than the white-light speed close to the Sun
(1774 km s$^{-1}$) but larger than the shock speed at 1 AU (1030 km
s$^{-1}$; see below), so the shock must be decelerated as the high
speed flow overtakes the preceding solar wind. Propagation of the
shock as well as the deceleration is calculated in \S 4.

Complementary plasma parameters from ACE are displayed in Figure~3;
the magnetic field is almost the same as measured at STEREO. The
ICME interval is identified by combining the enhanced helium/proton
density ratio and depressed proton temperature (as compared with the
normal temperature expected from the observed speed); the boundaries
also agree with the discontinuities in the density, bulk speed and
magnetic field. The resulting radial width (average speed times the
duration) is about 0.67 AU; the MC indicated by STEREO BDEs is about
the first half of the time interval. The preceding shock is also
apparent from the plasma parameters. The shock speed is about 1030
km s$^{-1}$ as calculated from the conservation of mass across the
shock, i.e., $v_s=(n_2v_2-n_1v_1)/(n_2-n_1)$, where $n_1=1.8$
cm$^{-3}$, $n_2=6.0$ cm$^{-3}$, $v_1=573$ km s$^{-1}$ and $v_2=896$
km s$^{-1}$ are average densities and speeds upstream (1) and
downstream (2) of the shock, respectively. As shown above, the shock
has to be decelerated when propagating from the Sun to the Earth. A
least squares fit of the plasma and magnetic field data across the
shock to the Rankine-Hugoniot relations \citep{vinas86} gives a
shock normal with elevation angle $\theta\simeq-12.7^{\circ}$ and
azimuthal angle $\phi\simeq308^{\circ}$ in RTN coordinates (with
${\bf R}$ pointing from the Sun to the spacecraft, ${\bf T}$
parallel to the solar equatorial plane and along the planet motion
direction, and ${\bf N}$ completing the right-handed system). The
shock normal makes an angle of about $56^{\circ}$ with the upstream
magnetic field, so the shock may be quasi-perpendicular.
\cite{liu06a} find that the sheath regions between fast ICMEs and
their preceding shocks are analogous to planetary magnetosheaths and
often characterized by plasma depletion layers and mirror-mode
waves. The proton density in the sheath of the current event first
increases and then decreases quickly close to the MC, very similar
to the case shown in their Figure~3. Magnetic fluctuations in the
sheath appear consistent with mirror-mode waves but note that large
depressions are due to current sheet crossings; a
quasi-perpendicular shock would heat the plasma preferentially in
the direction perpendicular to the field, so the plasma downstream
of the shock may be unstable to the mirror-mode instability
\citep[e.g.,][]{liu07}. There seems another ICME on December 17 as
can be seen from the low proton temperature and declining speed.
Interactions between the two events are likely present.

We reconstruct the MC structure using the Grad-Shafranov (GS)
technique which includes the thermal pressure and can give a cross
section without prescribing the geometry \citep[e.g.,][]{hau99,
hu02}. This method relies on the feature that the thermal pressure
and the axial magnetic field depend on the vector magnetic potential
only \citep{schindler73, sturrock94}, which has been validated by
observations from STEREO and ACE/Wind when these spacecraft are well
separated \citep{liu08}. We apply the method to the plasma and
magnetic field data between 22:48 UT on December 14 (the MC leading
edge) and 04:34 UT on December 15 (right before the current sheet
within the MC) when the magnetic field has the clearest rotation.
All the data used in the reconstruction are from the MC interior.
The reconstruction results are illustrated in Figure~4. The
recovered cross section (in a flux-rope frame with ${\bf x}$ along
the spacecraft trajectory and ${\bf z}$ in the direction of the
axial field) shows nested helical field lines, suggestive of a
flux-rope structure. The spacecraft (ACE and STEREO) seem to cross
the MC close to the axis with an impact parameter of 0.01 AU; the
maximum axial field is also very close to the leading edge ($x=0$).
The transverse fields along the spacecraft path indicate a
left-handed chirality. The reconstruction gives an axis elevation
angle $\theta\simeq -57^{\circ}$ and azimuthal angle $\theta\simeq
261^{\circ}$ in RTN coordinates, as shown in Figure~4 (right). Since
the axial field points southward and the field configuration is
left-handed, a spacecraft would see a field which is first most
negative and then becomes less negative as the MC passes the
spacecraft along the radial direction. These results are consistent
with observations (see Figures~2 and 3). We also apply the method to
a larger interval inside the MC (with the current sheet excluded)
and obtain a similar cross section and axis orientation. Note that
the axis orientation is oblique to the filament channel; similar
results are obtained for other cases \citep[e.g.,][]{wang06}. The MC
axis orientation depends on which part of the CME is observed in
situ, or the CME possibly rotates during the propagation in the
heliosphere.

\subsection{SEP Events}

Major SEP events are observed at 1 AU during the ICME passage. The
ICME structure as well as its effects on energetic particle
transport can also be inferred from particle measurements. Figure~5
shows the particle intensities measured by the Electron, Proton, and
Alpha Monitor (EPAM) of ACE and the Solar Electron and Proton
Telescope (SEPT) of STEREO A. Four particle enhancements are
evident. The timing with the flares and shocks indicates that the
first particle enhancement is associated with the injection from the
X3.4 flare (02:38 UT on December 13), the second one associated with
the ICME forward shock (14:38 UT on December 14), the third one with
an X1.5 flare (22:14 UT on December 14) which occurred in the same
active region (NOAA 10930; S06$^{\circ}$W46$^{\circ}$), and the
fourth one with the shock downstream of the ICME (17:23 UT on
December 16; see Figures~2 and 3). The X3.4 flare produced an
intense electron flux which declines during a long time period;
velocity dispersion is not clear in the electrons but present in the
protons. Note that the CME-driven shock should be producing
energetic particles throughout the interplanetary transit. It
continues to accelerate protons to $\sim$MeV energies at 1 AU but
appears to have a small effect on the electrons, probably because
the electron enhancement at the shock is masked by the large
preshock intensities. There is an apparent exclusion of the protons
from the ICME interior (see second panel), presumably screened off
by the strong fields within the ICME; the proton signature also
seems consistent with the BDE interval. Interestingly, there is an
intensity enhancement of the electrons within the ICME due to the
X1.5 flare; it is likely that the electrons stream along the field
line from the active region and are trapped inside the ICME
\citep[e.g.,][]{kahler91, larson97}, so the ICME may still be
magnetically connected to the Sun. These features are very similar
to the observations during the Halloween storm
\citep[e.g.,][]{malandraki05, mckibben05}.

The two bottom panels in Figure~5 show anisotropy information of the
particles provided by STEREO A/SEPT. SEPT has two separate
telescopes, one looking in the ecliptic plane along the nominal
Parker spiral field toward and away from the Sun and the other
looking vertical to the plane toward the south and north,
respectively \citep{muller07}. The lines in the two panels represent
the intensity differences between the two directions for each
telescope (i.e., differences between south and north and between the
two opposite directions along the Parker field); the intensity
difference is normalized by 4000 cm$^{-2}$ s$^{-1}$ sr$^{-1}$
MeV$^{-1}$ for the electrons and 1000 cm$^{-2}$ s$^{-1}$ sr$^{-1}$
MeV$^{-1}$ for the protons. The data before 18:10 UT on December 14
are discarded, because the SEPT doors were closed since the launch
of STEREO and were opened one by one from 17:32 UT to 18:10 UT on
December 14. The anisotropy information of the particles is thus not
available before 18:10 UT on December 14. The X1.5 flare produced
electrons and protons moving largely anti-sunward at 1 AU, but some
anti-sunward particles may be mirrored back by the enhanced magnetic
fields in the ICME/sheath and then propagate in the sunward
direction. A beam of particles accelerated at the second shock,
which may also be deflected by the first ICME, move sunward and
northward.

Further information about the electron behavior across the ICME is
provided by SWEA and the suprathermal electron instrument
\citep[STE;][]{lin08} on board STEREO as shown in Figure~6. The
electron energies range from $\sim$1 eV to 100 keV. The pitch angle
distribution of 247 eV electrons displayed in Figure~2 is measured
by SWEA in the spacecraft frame; BDE signatures are discernible from
$\sim$50 eV to $\sim$1.7 keV for this event. There seems a flux
decrease associated with the BDE interval; the high-energy electrons
from STE recover before the trailing edge of the BDE interval,
earlier than the low-energy electrons. The electron intensity
profiles are quantitatively different from what is shown in
Figure~5, which may be due to different looking directions of the
detectors and different connectivity of the spacecraft via the
magnetic field line to the Sun.

The above results suggest that the particle transport is largely
governed by the large-scale transient structures. The particles can
be deflected and constrained by ICMEs and reaccelerated by their
associated shocks, so modeling of the particle transport is
complicated by the presence of these structures. The energetic
particles, however, could be used to trace the ICME structure, which
can then be compared with plasma and magnetic field measurements.
More information on ion intensities, spectra, and composition
regarding the events can be found in \cite{mewaldt07} and
\cite{mulligan07}.

\subsection{ICME at Ulysses and Beyond}

Ulysses was at a distance 2.73 AU, latitude $-74.9^{\circ}$ and
longitude $123.3^{\circ}$ in the heliographic inertial system when
it observed a large shock at 17:02 UT on December 17. Ulysses was
about $117^{\circ}$ east and $74^{\circ}$ south of the Earth. The
large spacecraft separation provides a great opportunity to measure
the spatial extent of the CME-driven shock. Figure~7 displays the
Ulysses data for a 5-day interval. The shock is apparent from the
sharp increases in the plasma and magnetic field parameters. During
this time period, there are no clear ICME signatures such as
enhanced helium abundance, depressed proton temperature, and smooth
strong magnetic fields compared with the ambient solar wind upstream
of the shock. It is likely that the ICME is missed at Ulysses
whereas the shock is observed. The shock has a speed about 870 km
s$^{-1}$, smaller than the 1 AU speed (1030 km s$^{-1}$) but not
significantly. Given a speed difference larger than $740$ km
s$^{-1}$ between at the Sun and 1 AU, the primary deceleration of
the shock must occur within 1 AU, and further out the shock moves
with a roughly constant speed. Propagation of the shock from the Sun
to Ulysses is quantified in \S 4 by combining the coronagraph,
radio, and in situ data with an MHD model. The shock normal has an
elevation angle $\theta\simeq -38.3^{\circ}$ and azimuthal angle
$\phi\simeq 77.5^{\circ}$ (RTN), resulting from the Rankine-Hugoniot
calculations \citep{vinas86}. The angle between the shock normal and
the upstream magnetic field is about $68^{\circ}$, so the shock is
also quasi-perpendicular at Ulysses.

To show that Ulysses observed the same shock as ACE/STEREO, we
propagate the solar wind data outward from 1 AU using an MHD model
\citep{wang00}. The model has had success in connecting solar wind
observations at different spacecraft \citep[e.g.,][]{wang01,
richardson02, richardson05, richardson06, liu06b}. The model assumes
spherical symmetry (1D) since we have solar wind measurements at
only a single point. We use the solar wind parameters observed at 1
AU (50 days long around the ICME) as input to the model and
propagate the solar wind outward. Figure~8 shows the observed speeds
at 1 AU and Ulysses and the model-predicted speeds at certain
distances. Small streams smooth out due to stream interactions as
can be seen from the traces, but the large stream associated with
the shock persists out to Ulysses. The predicted arrival time of the
shock at Ulysses is only about 3.6 hr earlier than observed. The
time difference is negligible compared with the propagation time of
$\sim$75.1 hr from ACE to Ulysses. The ambient solar wind predicted
by the model is slower than observed at Ulysses, which is reasonable
since Ulysses is at the south pole ($74^{\circ}$ south of ACE).
Given the good stream alignment, we think that Ulysses and the
near-Earth spacecraft observed the same shock. Ulysses may be
observing the shock flank if the nose of the ICME is close to the
ecliptic plane. The successful model-data comparison also indicates
that the global shock surface is nearly spherical and the shock
speed variation from the shock nose to flank is small. It is
surprising that even at the south pole the shock is still observed;
spacecraft configured as above (i.e., one close to the solar
equatorial plane and the other at the pole) are rare while they
observe the same shock. Note that the longitudinal size of the shock
is also large with a lower limit of $117^{\circ}$ (the longitudinal
separation between the Earth and Ulysses).

The large size of the shock indicates that the global configuration
of the solar wind can be altered as the CME sweeps through the
heliosphere. We also propagate the solar wind to large distances
using the MHD model. The peak solar wind speeds quickly decrease as
the high-speed flow interacts with the ambient medium. They are
reduced to 630 km s$^{-1}$ at 10 AU and to 490 km s$^{-1}$ (close to
the ambient level) by 50 AU. Therefore, the high-speed streams would
not produce significant effects at large distances.

\section{CME/Shock Propagation}

Of particular interest for space weather forecasting is the
CME/shock propagation in the inner heliosphere. In the absence of
observations of plasma features, propagation of shocks within 1 AU
can be characterized by type II radio emissions
\citep[e.g.,][]{reiner07}. Type II radio bursts, typically drifting
downward in frequency, are remote signatures of a shock moving
through the heliosphere and driving plasma radiation near the plasma
frequency and/or its second harmonic \citep[e.g.,][]{nelson85,
cane87}. The frequency drift results from the decrease of the plasma
density as the shock propagates away from the Sun. The plasma
frequency, $f_p~({\rm kHz}) = 8.97\sqrt{n~({\rm cm^{-3}})}$, can be
converted to a heliocentric distance $r$ by assuming a density model
$n = n_0/r^2$
\begin{equation}
r~({\rm AU}) = \frac{8.97\sqrt{n_0~({\rm cm^{-3}})}}{f_p~({\rm
kHz})},
\end{equation}
where $n_0$ is the plasma density at 1 AU. The height-time profile
of shock propagation can then be obtained from the frequency drift.

Figure~9 displays the dynamic spectrum as well as soft X-ray flux
associated with the CME. An intense type III radio burst occurred at
about 02:25 UT on December 13 (Day 347), almost coincident with the
peak of the X-ray flux. Type III bursts are produced by
near-relativistic electrons escaping from the flaring site
\citep[e.g.,][]{lin73}, so they drift very rapidly in frequency and
appear as almost vertical features in the dynamic spectrum. Such an
intense type III burst often indicates a major CME \citep{reiner01}.
Note that many short-lived type III-like bursts are also seen
starting from 17:00 UT on December 13; they are known as type III
storms and presumably associated with a series of small electron
beams injected from the Sun. Diffuse type II emissions occur at the
fundamental and harmonic plasma frequencies and appear as slowly
drifting features. They start after the type III burst and seem
disrupted during the small flares around 13:00 UT on December 13
(see the X-ray flux). It is not clear whether the type II emissions
after 16:00 UT on December 13 are at the fundamental or harmonic of
the plasma frequency; the broad band may result from merging of the
two branches. Apparently it is difficult to measure the frequency
drift from individual frequencies associated with the type II
bursts. An overall fit combined with in situ measurements at 1 AU
would give a more accurate estimate for the height-time profile.

We employ a kinematic model to characterize the CME/shock
propagation, similar to the approach of \cite{gopalswamy01} and
\cite{reiner07}. The shock is assumed to start with an initial speed
$v_0$ and a constant deceleration $a$ lasting for a time period
$t_1$, and thereafter it moves with a constant speed $v_s$. The
shock speed $v_s$ and transit time $t_T$ are known from 1 AU
measurements, leaving only two free parameters in the model ($a$ and
$t_1$). At a time $t$, the distance of the shock can be expressed as
\begin{equation}
r=\left\lbrace
\begin{array}{ll}
d + v_s(t-t_T) + a(\frac{1}{2}t^2+\frac{1}{2}t_1^2 -t_1t)~~~~~t<t_1\\
d + v_s(t-t_T)~~~~~~~~~~~~~~~~~~~~~~~~~~~~~~~\,t\geq t_1
\end{array} \right.
\end{equation}
where $d=1$ AU and $v_0=v_s - at_1$. The trace of the fundamental
branch of the type II bursts is singled out using an interactive
program and shown in Figure~10; the selected frequencies are
converted to heliocentric distances using equation~(1). We adjust
the density scale factor $n_0$ to obtain a best fit of the frequency
drift; a value of $n_0\simeq 13$ cm$^{-3}$ gives a height-time
profile that simultaneously matches the radio data and the shock
parameters at 1 AU. The density model describes the average radial
variation of the ambient density, so the scale factor is not
necessarily the observed plasma density upstream of the shock at 1
AU. Two curves corresponding to the emissions at the fundamental and
harmonic plasma frequencies are obtained from the best fit, as shown
in Figure~9. The fit is forced to be consistent with the overall
trend of the frequency-drifting bands; discrepancies are seen at
some times due to irregularities of the type II emissions. Note that
the best fit yields the radial kinematic parameters of the CME/shock
propagation with projection effects minimized. The radial velocity
of the shock near the Sun given by the best fit is $v_0\simeq2212$
km s$^{-1}$, larger than the measured CME speed projected onto the
sky (1774 km s$^{-1}$). The deceleration is about $-34.7$ m
s$^{-2}$, lasting for $\sim$9.5 hr which corresponds to a distance
of about 0.36 AU. Thereafter the shock moves with a constant speed
1030 km s$^{-1}$ as measured at 1 AU.

In order to show how well the CME/shock is tracked by the fit, we
extend the curve to the distance of Ulysses and plot in Figure~10
the CME locations measured by LASCO (see the LASCO CME catalogue at
\url{http://cdaw.gsfc.nasa.gov}), the MHD model output of the shock
arrival times every 0.2 AU between 1 - 2.6 AU, and the shock arrival
time at Ulysses. The fit agrees with the LASCO data, the MHD model
output at different distances and finally the Ulysses measurement.
Note that we only use the type II frequency drift and the 1 AU shock
parameters to obtain the height-time profile. Even at large
distances the shock is still tracked remarkably well by the fit. The
agreement verifies the kinematic model for the CME/shock
propagation; a value of 2212 km s$^{-1}$ should be a good estimate
of the CME radial velocity near the Sun. The separation between the
shock and CME should be very small near the Sun but is not
negligible at 1 AU (see Figures~1 and 3).

These results present an important technique for space weather
forecasting, especially when in situ measurements closer to the Sun
are available (say, from the Solar Orbiter and Sentinels). In situ
data closer to the Sun can be propagated to 1 AU by an MHD model;
further constraints on the height-time profile are provided by the
frequency drift of type II emissions. The advantage of this method
is that the shock can be tracked continuously from the Sun all the
way to 1 AU; the arrival time of CME-driven shocks at the Earth can
be predicted with an accuracy less than a few hours $\sim$days
before they reach the Earth. Implementation of the method,
specifically combining MHD propagation of the solar wind with type
II frequency drift, is expected to be a routine possibility in the
future when in situ data are available from the Solar Orbiter and
Sentinels.

\section{Summary and Discussion}

We have investigated the evolution and propagation of the 2006
December 13 CME combining remote sensing and in situ measurements
with modeling efforts. A comprehensive view of the CME is made
possible by coordinated EUV, coronagraph, radio, particle and in
situ plasma and magnetic field observations provided by a fleet of
spacecraft including SOHO, STEREO, ACE, Wind and Ulysses.

The CME is accompanied by an X3.4 solar flare, EUV dimmings and EIT
waves. It had a speed about 1774 km s$^{-1}$ near the Sun and
produced SEP events, radio bursts, an interplanetary shock, and a
large ICME embedded with an MC which gave rise to a major
geomagnetic storm. The speed of the CME-driven shock is about 1030
km s$^{-1}$ at 1 AU, suggestive of a significant deceleration
between the Sun and 1 AU. Reconstruction of the MC with the GS
method indicates a flux-rope structure with an axis orientation
oblique to the flare ribbons. We observe major SEP events at 1 AU,
whose intensities and anisotropies are used to investigate the ICME
structure. The ICME is still magnetically connected to the Sun, as
indicated by the electron enhancement due to the X1.5 flare within
the ICME. Particle deflection and exclusion by the ICME suggest that
the energetic particle transport is largely dominated by the
transient structures.

The CME-driven shock is also observed at Ulysses while the ICME
seems missed. Ulysses was 74$^{\circ}$ south and $117^{\circ}$ east
of the Earth, indicative of a surprisingly large angular extent of
the shock. The shock speed is about 870 km s$^{-1}$, comparable to
its 1 AU counterpart. An MHD model using the 1 AU data as input
successfully predicts the shock arrival time at Ulysses with a
deviation of only 3.6 hr, substantially smaller than the propagation
time 75.1 hr from ACE to Ulysses. The model results also show that
the peak solar wind speeds quickly decrease at large distances.
Consequently, the CME/shock would not cause large effects in the
outer heliosphere.

To the best of our knowledge, this may be the largest CME-driven
shock ever detected in the space era. Ulysses, launched in 1991, is
the only spacecraft that can explore the solar wind conditions at
high latitudes. A survey of ICMEs from observations of near-Earth
spacecraft and Ulysses shows that these spacecraft are generally
separated within 40$^{\circ}$ in latitude when they observe the same
CME-driven shock \citep{liu05, liu06b}. \cite{reisenfeld03} report
an ICME as well as a preceding shock observed at both ACE and
Ulysses with a latitudinal separation 73$^{\circ}$ (comparable to
the present one), but the longitudinal separation of the spacecraft
is only 64$^{\circ}$, much smaller than the current case. At the
time of the Bastille Day event in 2000, Ulysses was 65$^{\circ}$
south and 116$^{\circ}$ east of the Earth, comparable to but smaller
than the present spacecraft separation; the shock as well as the
ICME, however, did not reach Ulysses \citep{zhang03}. During the
record-breaking Halloween storm in 2003, the spacecraft that
observed the preceding shock were all at low latitudes
\citep{richardson05}. Other documented events in the last 150 years
either occurred before the space era or were associated with
spacecraft separations smaller than the current one
\citep{burlaga95, cliver04, gopalswamy05, richardson02,
richardson06}.

Tracking the interplanetary transport of CME-driven shocks (as well
as measuring their global scale) is of critical importance for
solar, heliospheric, and magnetospheric studies and space weather
forecasting. We draw particular attention to the CME/shock
propagation combining coronagraph images, type II bursts, in situ
measurements and the MHD model. The height-time profile is deduced
from the frequency drift of the type II bands and the shock
parameters measured at 1 AU assuming a kinematic model;
uncertainties in the frequency drift are minimized by the
constraints from 1 AU data. The shock is tracked remarkably well by
the height-time curve, as cross verified by LASCO data, the MHD
model output at different distances and Ulysses observations. The
CME/shock has a radial speed of 2212 km s$^{-1}$ near the Sun; the
effective deceleration is about $-$34.7 m s$^{-2}$ and lasts for 9.5
hr corresponding to a transit distance of 0.36 AU. These results
demonstrate that a shock can be tracked from the Sun all the way to
1 AU (and larger distances) by combining MHD propagation of the
solar wind and type II emissions, a crucial technique to predict the
shock arrival time at the Earth with small ambiguities especially
when in situ measurements closer to the Sun are available from the
Solar Orbiter and Sentinels.

\acknowledgments The research was supported by the STEREO project
under grant NAS5-03131. We acknowledge the use of SOHO, GOES, ACE,
Wind and Ulysses data and CME parameters from the LASCO CME
catalogue maintained by NASA and the Catholic University of America
in cooperation with NRL. We thank C. T. Russell for helping maintain
the STEREO/MAG data, and are grateful to the referee for his/her
helpful suggestions. This work was also supported in part by grant
NNSFC 40621003.

\clearpage

\begin{figure}
\centerline{\includegraphics[width=14pc]{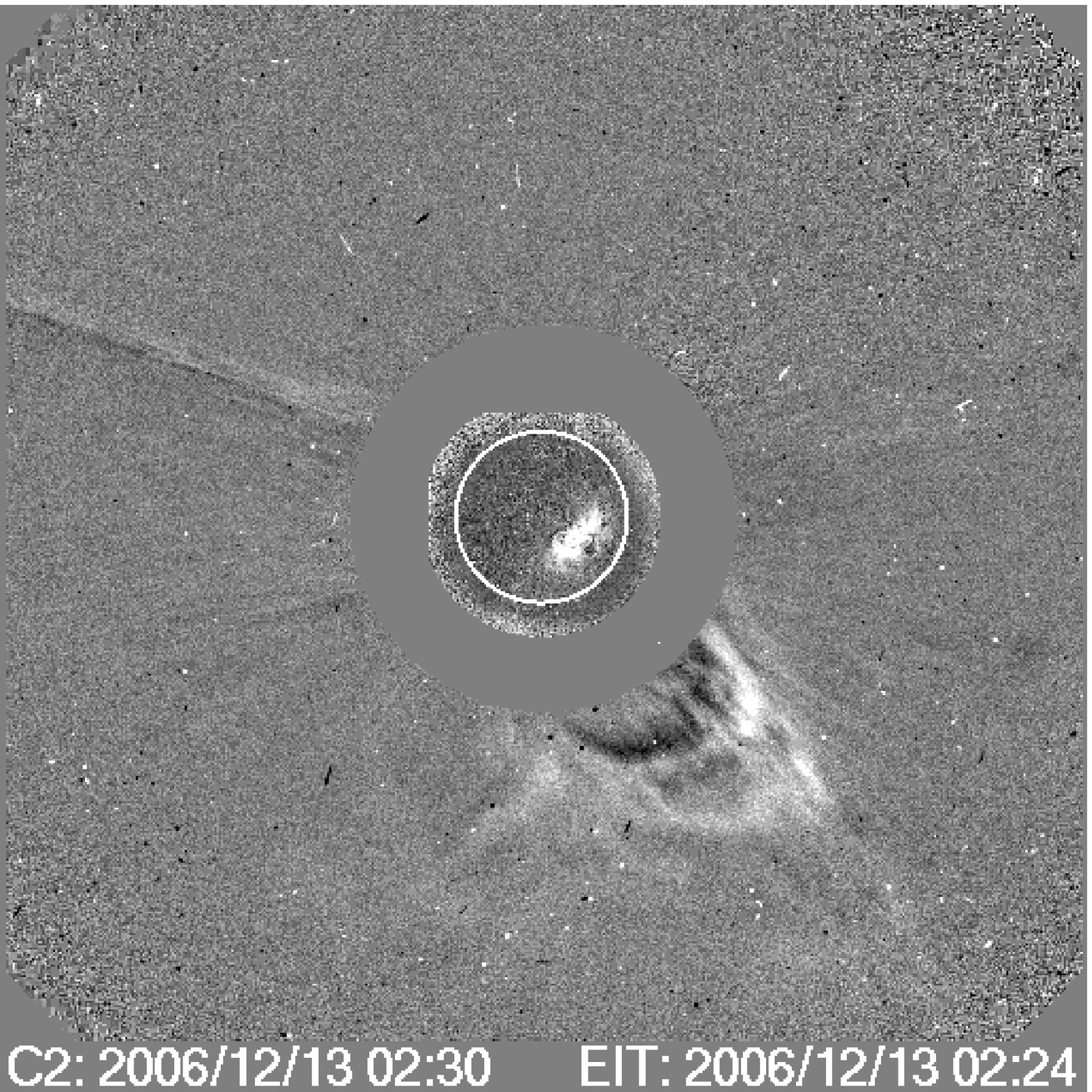}
\includegraphics[width=14pc]{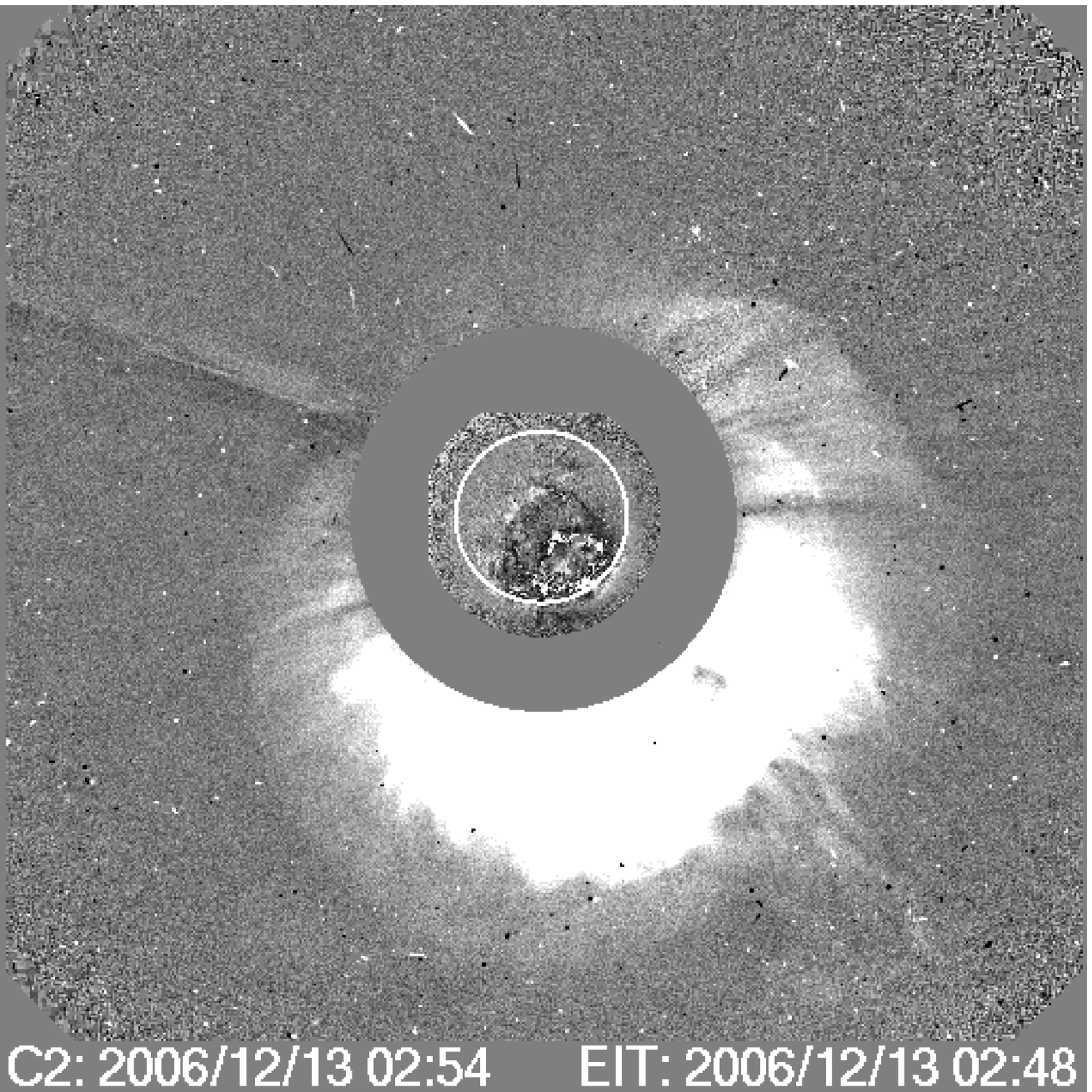}
\includegraphics[width=14pc]{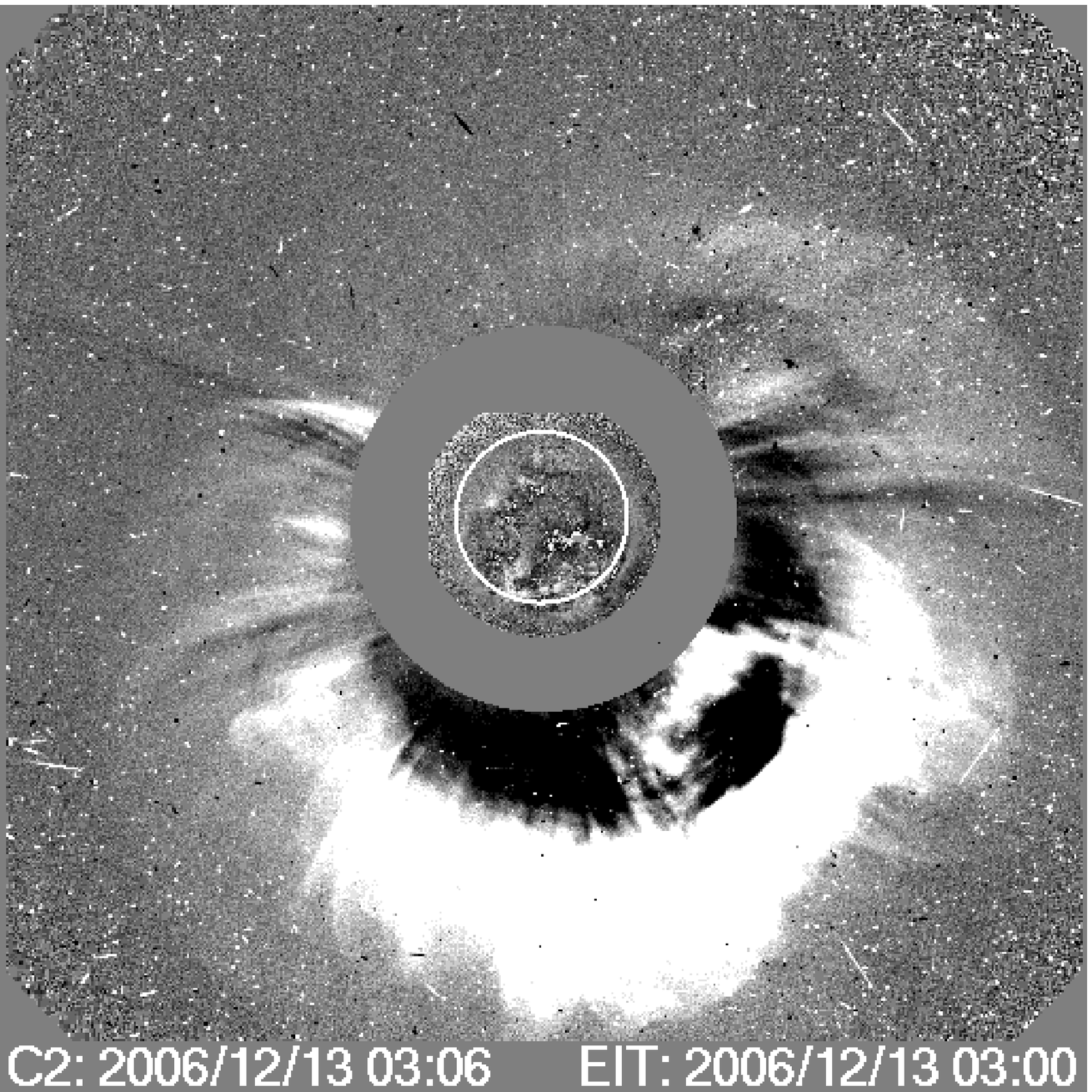}}
\caption{Difference images of the CME and source region at different
times. Filled in the circle are EIT difference images at 195 \AA. A
transition layer is visible around the CME front, indicating the
existence of a shock (middle and right panels). Adapted from the
LASCO CME catalogue at http://cdaw.gsfc.nasa.gov.}
\end{figure}

\clearpage

\begin{figure}
\epsscale{0.75} \plotone{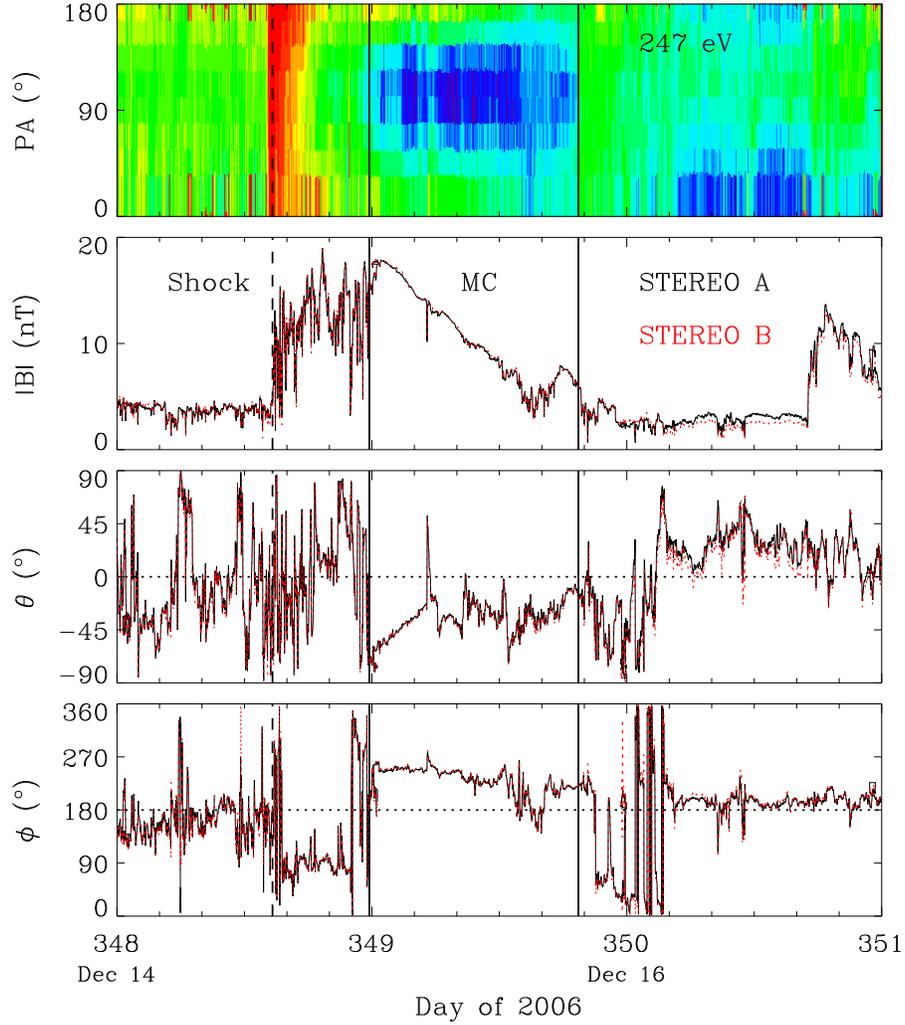} \caption{Pitch angle (PA)
distribution of 247 eV electrons measured by STEREO A, and magnetic
field strength, field elevation and azimuthal angles in RTN
coordinates measured by STEREO A (black) and B (red) across the MC
(bracketed by the two vertical lines). The dashed line denotes the
arrival time of the MC-driven shock. The color shading indicates
values of the electron flux (descending from red to blue).}
\end{figure}

\clearpage

\begin{figure}
\epsscale{0.80} \plotone{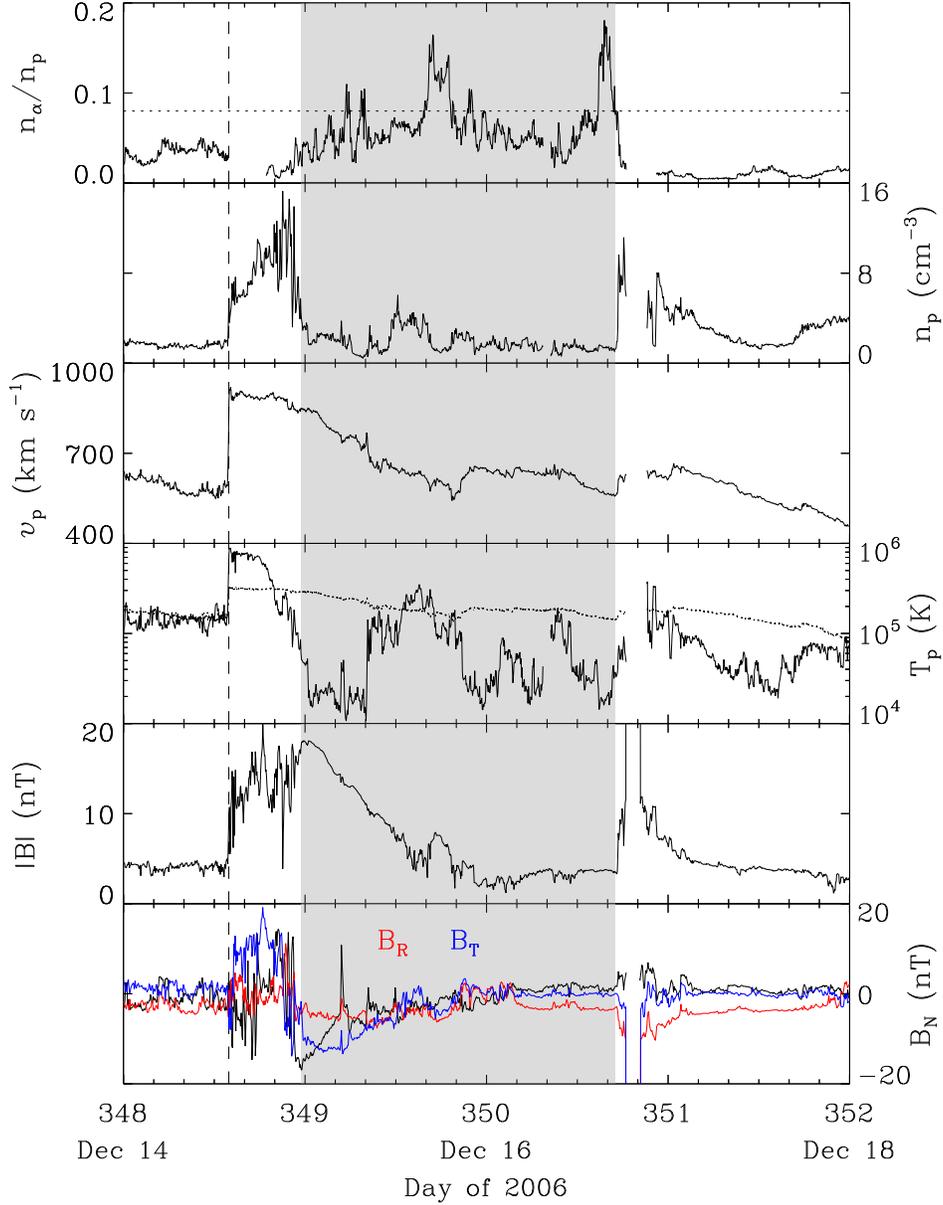} \caption{Solar wind plasma and
magnetic field parameters across the ICME (shaded region) observed
at ACE. From top to bottom, the panels show the alpha/proton density
ratio, proton density, bulk speed, proton temperature, magnetic
field strength and components in RTN coordinates. The dotted lines
denote the 8\% level of the density ratio (first panel) and the
expected proton temperature (fourth panel), respectively. The
arrival time of the shock is marked by the vertical dashed line.}
\end{figure}

\clearpage

\begin{figure}
\centerline{\includegraphics[width=24pc]{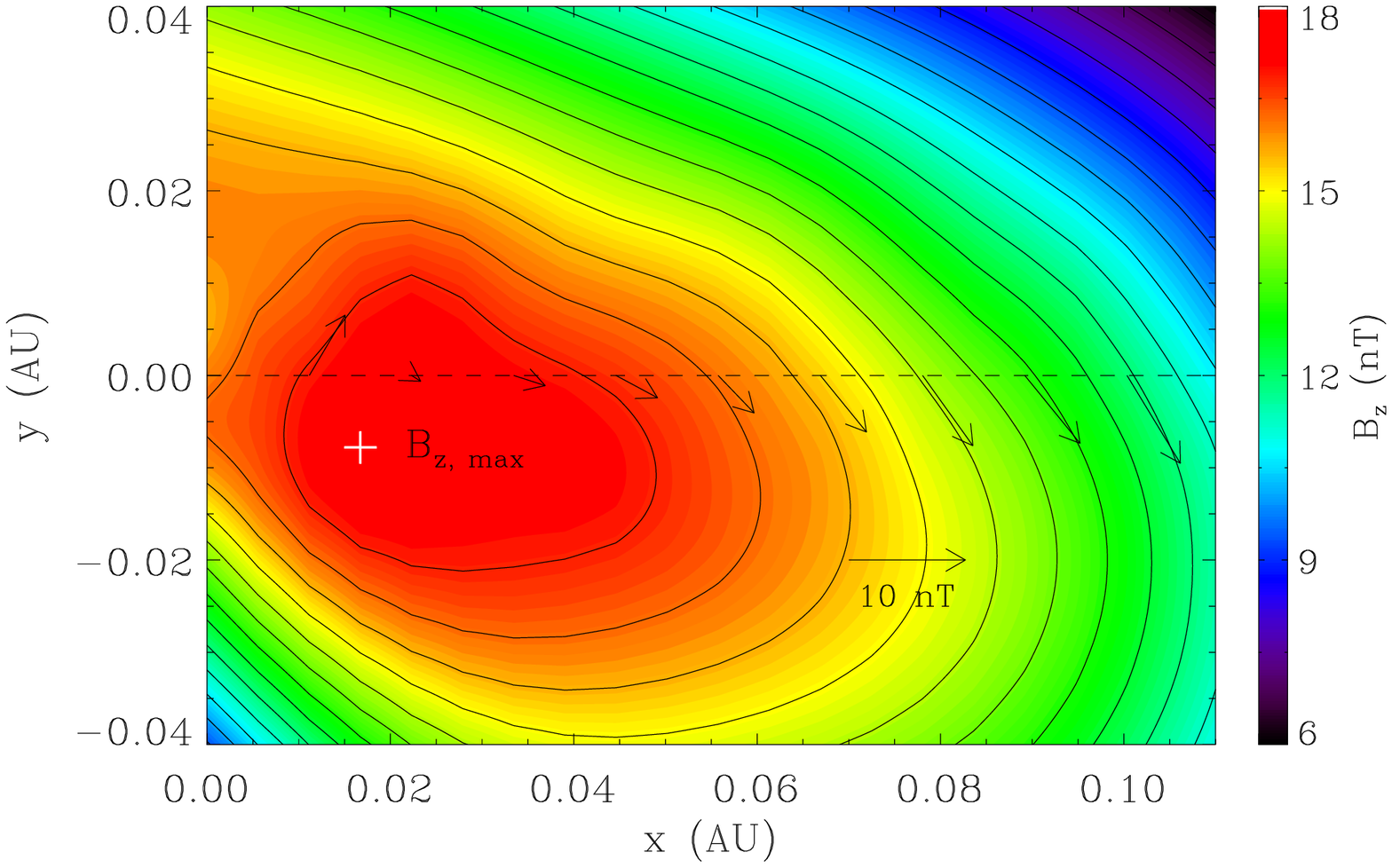}
\includegraphics[width=15pc]{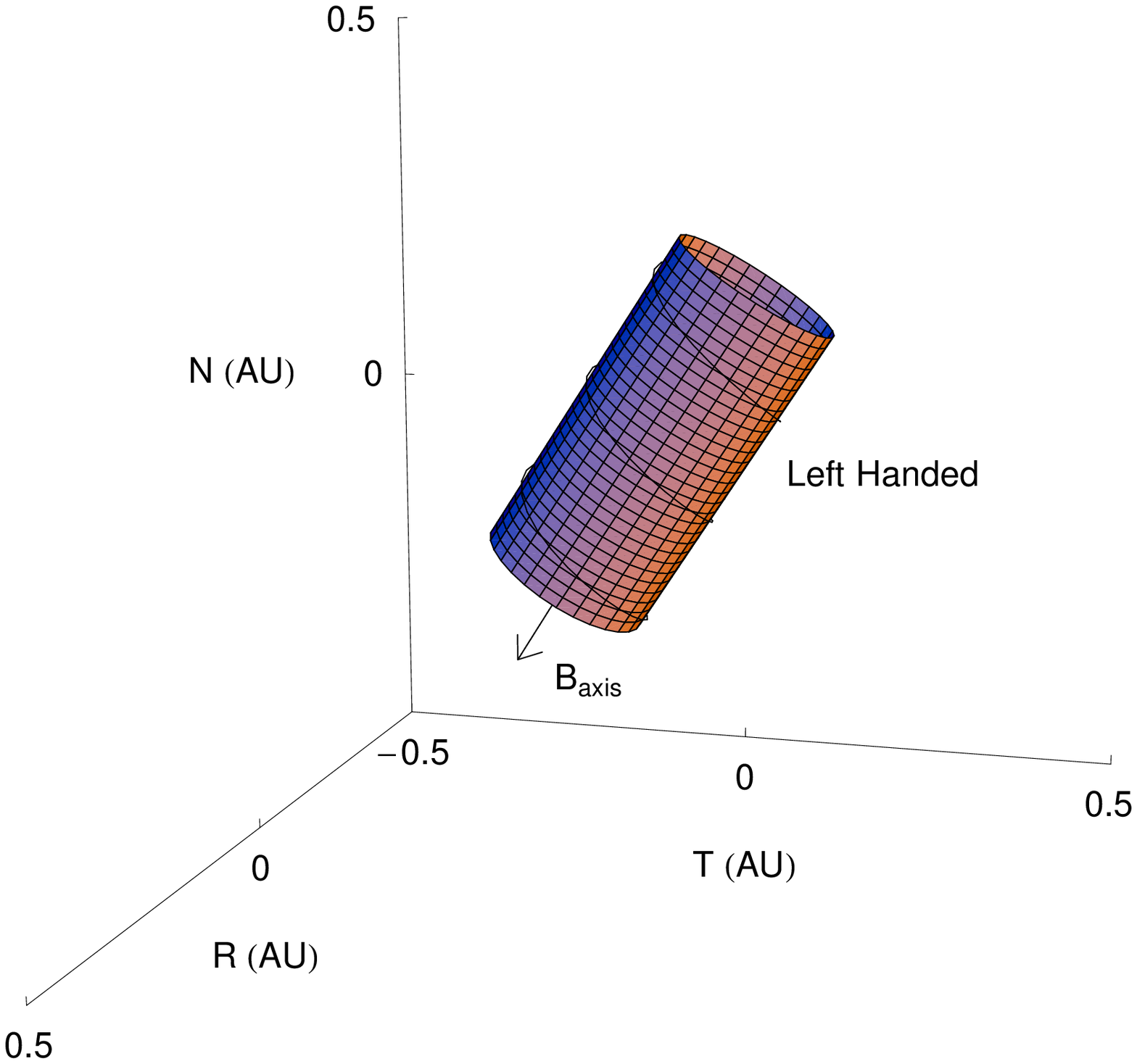}}
\caption{Left: Reconstructed cross section of the MC. Black contours
show the distribution of the vector potential and the color shading
indicates the value of the axial field. The dashed line marks the
trajectory of the spacecraft. The arrows denote the direction and
magnitude of the observed magnetic fields projected onto the cross
section. The location of the maximum axial field is indicated by the
plus sign. Right: An idealized schematic diagram of the MC
approximated as a cylindrical flux rope in RTN coordinates with the
arrow and helical line indicating the field orientation.}
\end{figure}

\clearpage

\begin{figure}
\epsscale{0.80} \plotone{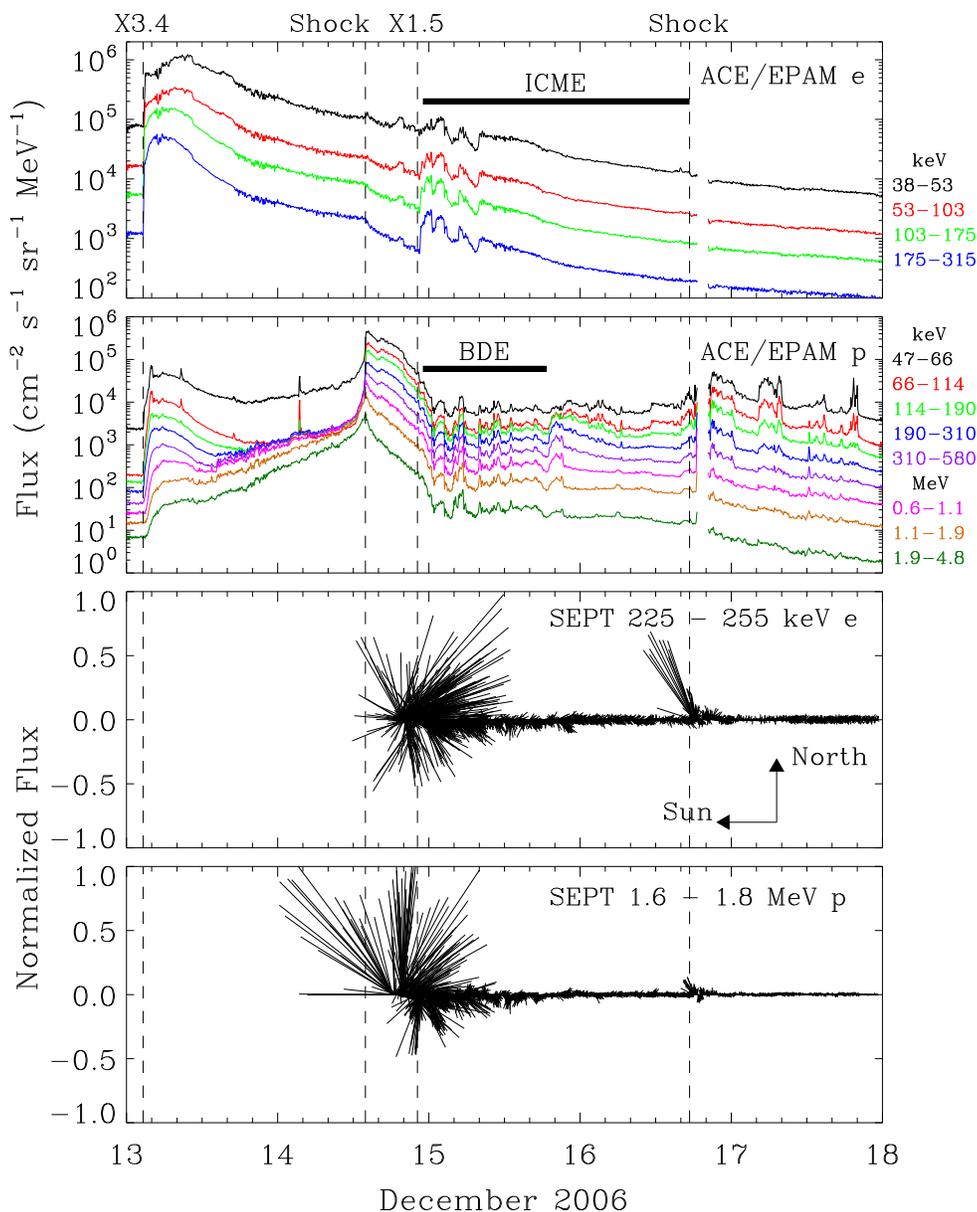} \caption{Intensities of electrons
(top panel) and protons (second panel) at different energy channels
measured by ACE/EPAM and anisotropies of electrons (third panel) and
protons (bottom panel) observed by STEREO A/SEPT. The times of the
flares and shocks are marked by the vertical dashed lines. The ICME
and BDE intervals are indicated by the horizontal bars. The solid
lines in the two bottom panels denote the normalized intensity
differences along the directions defined by the arrows (third
panel).}
\end{figure}

\clearpage

\begin{figure}
\epsscale{0.80} \plotone{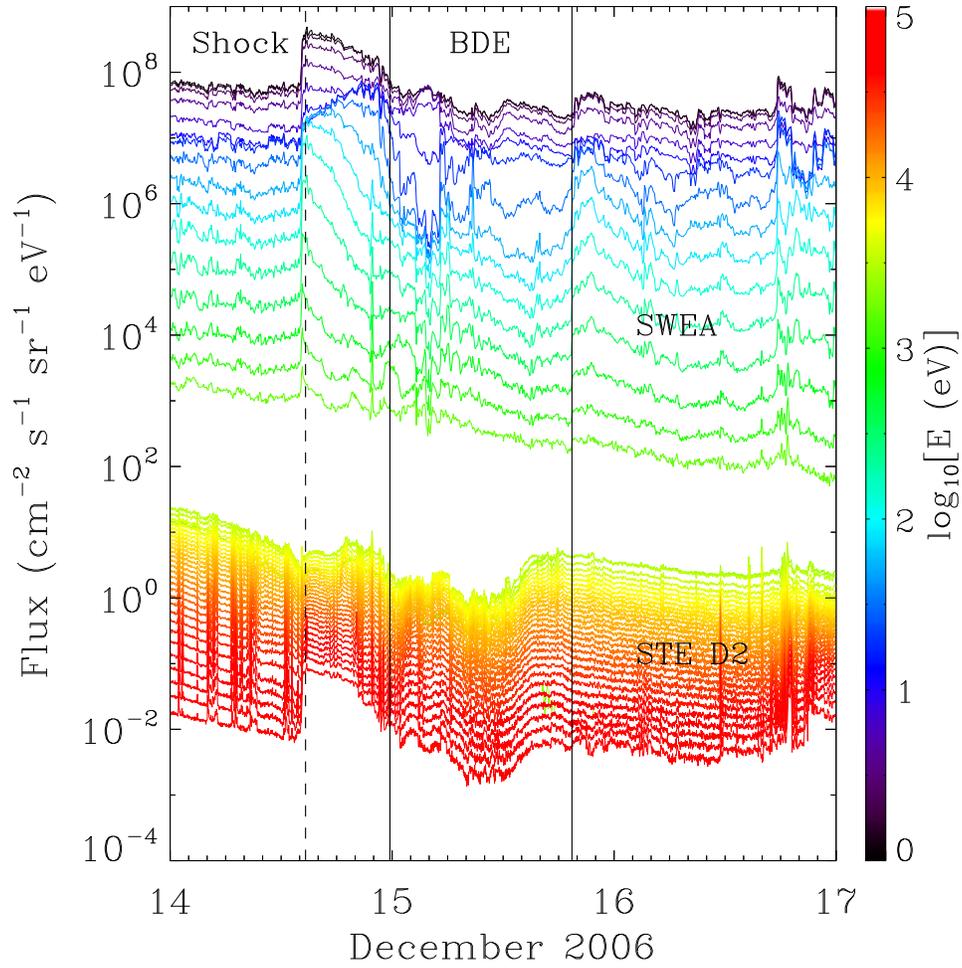} \caption{Electron intensities
across the BDE interval (between the solid lines) measured by SWEA
and STE D2 (one of the detectors looking away from the Sun) aboard
STEREO B. The color scale shows the electron energies. The shock
arrival time is indicated by the dashed line.}
\end{figure}

\clearpage

\begin{figure}
\epsscale{0.80} \plotone{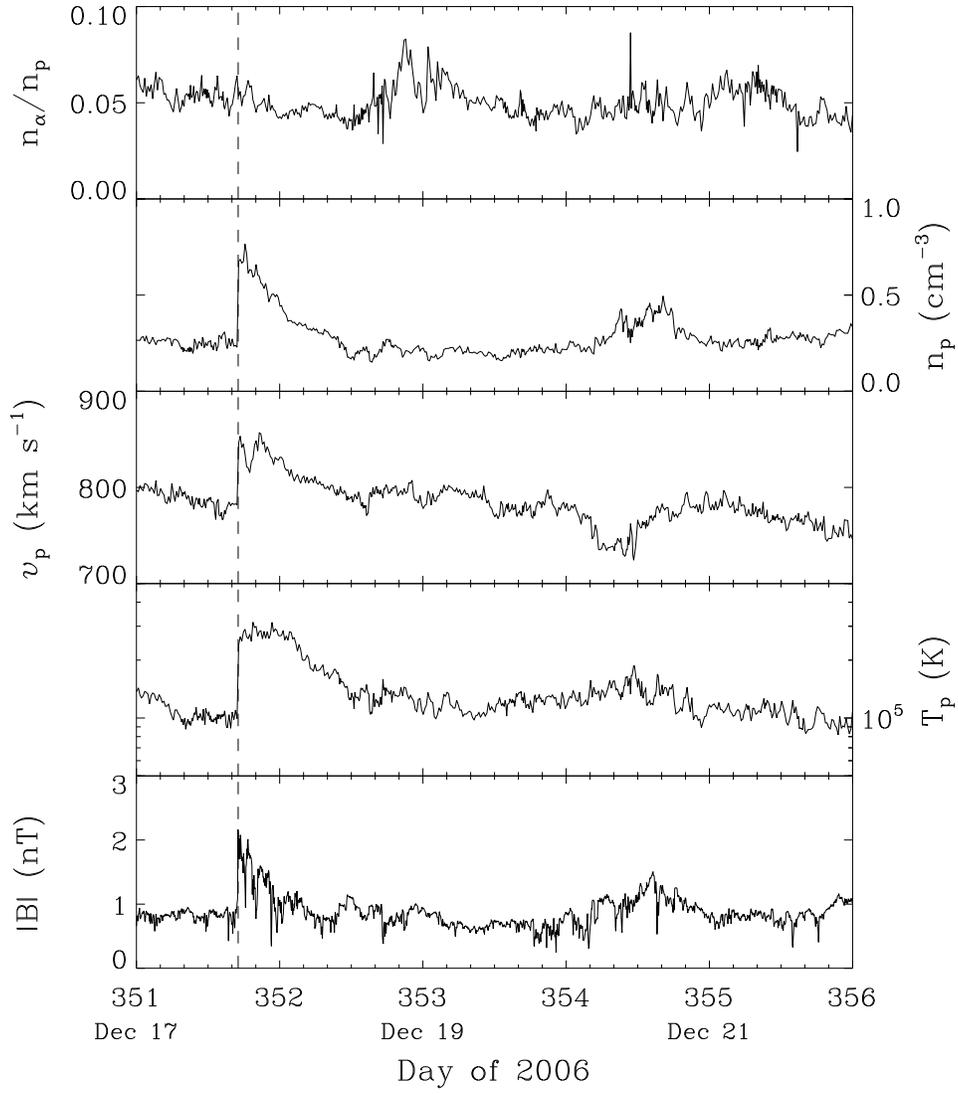} \caption{Same format as Figure~3,
but for the measurements at Ulysses.}
\end{figure}

\clearpage

\begin{figure}
\epsscale{0.80} \plotone{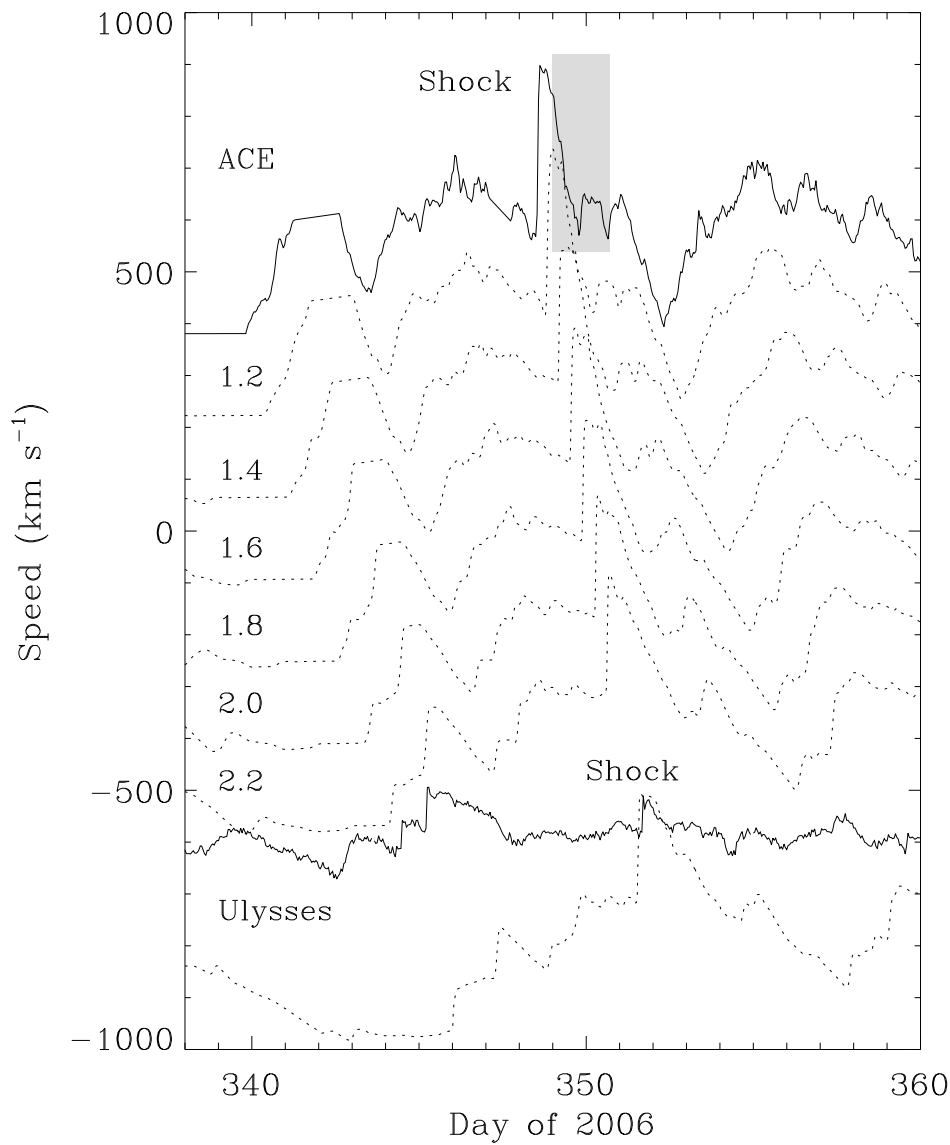} \caption{Evolution of solar wind
speeds from ACE to Ulysses via the MHD model. The shaded region
represents the ICME interval at ACE. The upper and lower solid lines
show the solar wind speeds observed at ACE and Ulysses. The dotted
lines denote the predicted speeds at the distances (in AU) marked by
the numbers; each curve is decreased by 160 km s$^{-1}$ with respect
to the previous one so that the individual profiles are discernable.
The speed profiles at Ulysses (both observed and predicted) are
shifted downward by 1360 km s$^{-1}$ from the 1 AU speeds.}
\end{figure}

\clearpage

\begin{figure}
\epsscale{1.00} \plotone{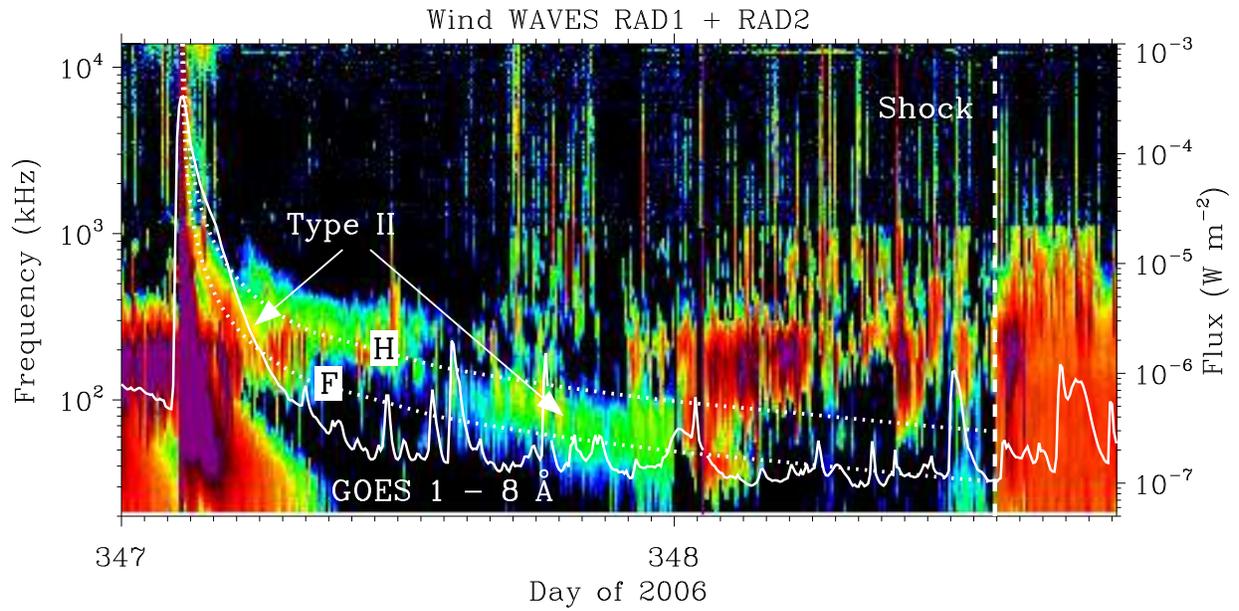} \caption{Dynamic spectrum (colors)
from Wind/WAVES and X-ray flux (solid line) from GOES 12. The dashed
vertical line indicates the arrival time of the preceding shock at 1
AU, and the dotted lines represent the best fits of the frequency
drift of the fundamental (F) and harmonic (H) type II bursts.}
\end{figure}

\clearpage

\begin{figure}
\epsscale{0.80} \plotone{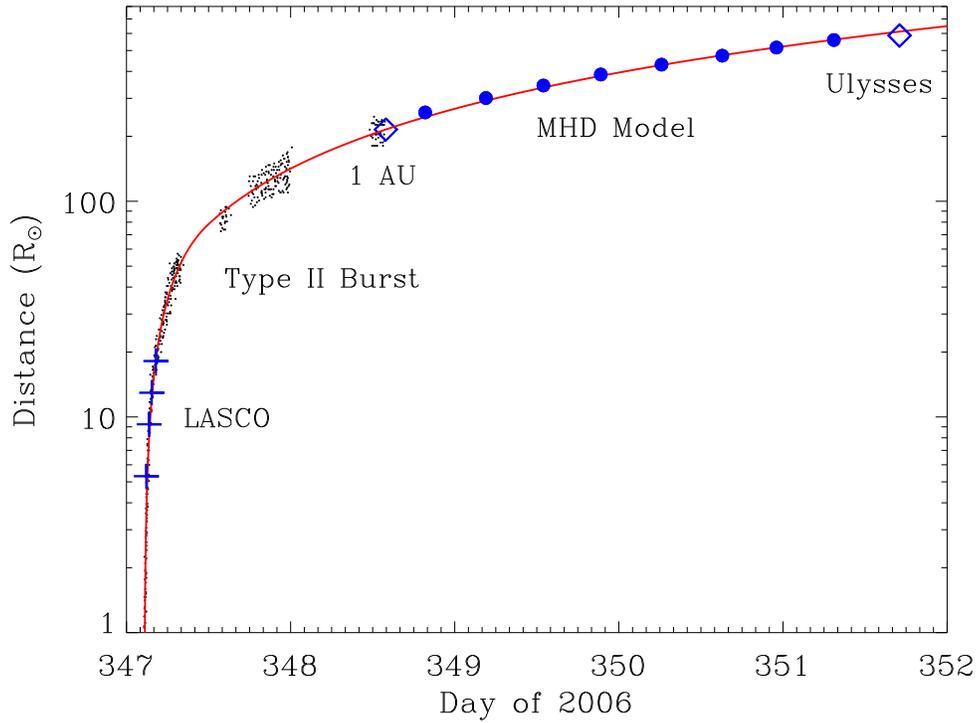} \caption{Height-time profile
(solid line) of shock propagation determined from the frequency
drift of the type II bands (dots) and shock parameters measured at 1
AU ($R_{\odot}$ being the solar radius). Pluses denote the LASCO
data. Diamonds indicate the shock arrival times at 1 AU and Ulysses.
Between 1 AU and Ulysses are the shock arrival times (filled
circles) at [1.2, 1.4, 1.6, 1.8, 2.0, 2.2, 2.4, 2.6] AU predicted by
the MHD model.}
\end{figure}

\end{document}